\documentclass[10pt,twocolumn]{revtex4-2} 
\usepackage[dvips]{color}
\usepackage{epsfig}
\usepackage{amssymb}
\usepackage{latexsym}
\usepackage{graphicx}
\begin{document}

\title{Application of Quaternion Neural Network to\\
                         Time Reversal Based Nonlinear Elastic Wave Spectroscopy}%
\author
{ Sadataka Furui}\affiliation{Faculty of Science and Engineering, Teikyo University, Utsunomiya, 320 Japan }
 \email{furui@umb.teikyo-u.ac.jp}
\author
{ Serge Dos Santos}\affiliation{INSA Centre\, Val de Loire,  Blois,
Inserm U1253, Universit\'e de Tours, Imagerie et Cerveau, imaging and brain : iBrain, France }
\email{serge.dossantos@insa-cvl.fr} 

\date{\today }

\begin{abstract}
Identification of crack positions or anomalies in materials using the time reversal based nonlinear elastic wave spectroscopy
 (TR-NEWS) is an established method. We propose a system using transducers which emit forward propagating solitonic wave and 
time-reversed propagating solitonic wave produced by memristors placed on a side of a complex medium, scattered by cracks in the 
material and received by receivers which are placed on the opposite side of the complex medium.

By minimizing the difference of the scattered forward propagating wave and the scattered TR wave, we get information of the 
position of the crack by using neural network techniques. Route of the solitons are expressed by 2 dimensional projective 
quaternion functions, and parameters for getting the optimal route from signals are expected to be reduced.

We consider the wave is expressed by a soliton which is conformal, and discuss symmetry protected topological impurities and 
gravitational effects using the Atiyah-Patodi-Singer's index theorem. 
\end{abstract}
\maketitle
\section{Introduction}
Non destructive testing (NDT) in solids was performed by Walsh\cite{Walsh65}, Fink et al.\cite{TRF94,Fink99}, McCall and Guyer
\cite{MCG94} by using phonetic waves. Sutin et al. \cite{STCJ04} used time reversed (TR) waves. Dos Santos et al.\cite{DSCSS06} improved his method to Time reversal based nonlinear elastic wave spectroscopy (TR-NEWS).

In non destructive testing (NDT), time reversal (TR) based nonlinear elastic wave spectroscopy (NEWS)
\cite{GCDSBM07,GDSBMC08,DosSantos20} is an efficient method to detect scatterers of ultra high frequency phonons in materials
\cite{SMJCBDS14,PM22,FM21,BRVW18,PWG21}.

At present, position of scatterers are defined by manually looking for an angle of the receiver relative to the transducer where 
interference of original waves and TR waves produce a peak. If there are several transducers which emit the wave pairs, and several receivers that measure convolutions of various pair waves, one can imagine getting interference patterns of beam pairs from different 
transducers. In \cite{FM21}, a method of putting a transducer at a corner of quadrate, placing receivers on the quadrate at regular intervals, and measuring the sum of received TR waves, which is called raster TR method is proposed. We consider in this work the standard TR method.

The information one obtains is large, but using techniques of neural networks\cite{PWG21,Aggarwal18,RLM22,Mathematica12}, it may be possible to detect scattering positions. An aim of this paper is to present a technique for this purpose.

Propagation of elastic phonetic waves with the direction ${\bf [110]}$ described by
\begin{equation}
w=w_0 exp[\sqrt{-1}(K_x x+K_y y-\omega t)],
\end{equation}
where $w$ is the shift of a particle in the material, $K_x, K_y$ are wave vector $2\pi/\lambda$.\cite{Kittel53}. The angular frequency $\omega$ depends on whether shift is longitudinal or transverse.

Direct determination of solitonic waves in waveguide was discussed by several authors\cite{KK78,Efimov79}.  Samsonov et al. \cite{SSB17} showed that strain soliton propagations in waveguides is measurable.

Inelastic scattering of phonons with the wave vector $\bf K^1$ and $\bf K^2$ can construct a phonon $\bf K^3$ through nonlinear effects. Interaction of phonons with lattices of materials yields
\begin{equation}
\sum_n exp[\sqrt{-1}({\bf K^3}-{\bf K^1}-{\bf K^2})\cdot {\bf r}_n],
\end{equation}
where ${\bf r}_n$ is the coordinate of the lattice vertex, and scattering conditions
\begin{equation}
{\bf K^3}={\bf K^1}+{\bf K^2}, \quad {\rm or} \quad {\bf K^3}={\bf K^1}+{\bf K^2}+{\bf G},
\end{equation}
where $\bf G$ is the inverse lattice vector, emerge.


Approximate solutions of nonlinear acoustic wave equation in materials with axial symmetry similar to the equation of phonons with the direction ${\bf [110]}$ which is equivalent to the Khokhlov-Zabolotskaya(KhZa) equation were considered in\cite{ZK69,LR84,LR92}. (In order to evade confusion with the Knizhnik-Zamolchikov(KZ) equation\cite{Wiki19}, which is relevant to the $SU(2)$ Wess-Zumino-Witten model, we use the abbreviation differnt from that of \cite{DSBM04}.)  Solitonic propagation for NDT applications has been the subject of several studies \cite{Walsh65,Fink99,GDSBMC08,BRVW18}.

The evolution of a KhZa soliton is given by a solution of
\begin{equation}
\frac{\partial u}{\partial x}=-\frac{\epsilon}{{c_0}^2}u \frac{\partial u}{\partial \tau}=-\frac{2}{2r}\frac{\partial}{\partial r}(r v),\quad \frac{\partial v}{\partial\tau}+c_0\frac{\partial u}{\partial r}=0.
\end{equation}
Lapidus and Rudenko\cite{LR84} considered the particle velocity along the beam direction $u$, and across the beam direction $v$,  using dimensionless variables 
\begin{eqnarray}
&&V=u/u_0, U=v\cdot (2 l_d/u_0 a), \nonumber\\
&&\theta=\omega(t-x/c_0),  z=x/l_s, R=r/a, 
\end{eqnarray}
where $c_0$ is the sound velocity, $x$ and $r$ are the axial and the transverse coordinates,
$l_d=\omega a^2/2c_0$ is the diffraction length, $l_s=c_0^2/\epsilon \omega u_0$, $\epsilon$ is the nonlinearity parameter, $N=l_s/l_d$, $u_0,\omega,a$ are characteristic values of the amplitude, frequency, and beam radius.

The equation is transformed to
\begin{eqnarray}
&&\frac{\partial V}{\partial z}-V\frac{\partial V}{\partial\theta}=-\frac{N}{4}\frac{1}{R}\frac{\partial}{\partial R}(RU)\nonumber\\
&&\frac{\partial U}{\partial\theta}+\frac{\partial V}{\partial R}=0.
\end{eqnarray} 

An exact solution of the KhZa equation is obtained by rewriting the equation 
\begin{eqnarray}
&&\frac{\partial^2 V}{\partial T\partial z}-\frac{N}{4}(\frac{\partial ^2 V}{\partial R^2}+\frac{1}{R}\frac{\partial V}{\partial R})=0,\nonumber\\
&&\frac{f'(z)}{f(z)}V\frac{\partial V}{\partial T}+\frac{\partial V}{\partial z}\frac{\partial V}{\partial T}-\frac{N}{4}(\frac{\partial V}{\partial R})^2=0,
\end{eqnarray}
where $T=\theta+z V$ in the case of a plane wave propagation.

A simple solution of equation the KhZa equation given in \cite{LR92} is
\begin{eqnarray}
V&=&\frac{C}{f(z)} exp[\sqrt{-1}T-R^2\frac{1-\sqrt{-1}b}{f(z)}],
\nonumber\\
&&f(z)=1-\sqrt{-1}N z(1-\sqrt{-1}b),
\end{eqnarray}
where $C$ and $b$ are constant. The solution can be checked by using Mathematica\cite{Mathematica12}.

The function $V$ is complex, but the real part describes the propagation of a focused harmonic wave with a Gaussian transverse distribution.  $Re V$ indicates the $\hat u-$component of the wave front coordinate,  and $Im V$ indicates the $\hat v-$component of the wave front, that propagates on the $(\hat u,\hat v)$ plane. 

In the limit of $z=x/l_s=0$,
\begin{eqnarray}
V&=&C exp[\sqrt{-1}\omega(t-x/c_0) ] exp[-r^2(1-b N)]\nonumber\\
&&\times(\cos(-b r^2)+\sqrt{-1}\sin(-b r^2)).
\end{eqnarray}

In the analysis of propagation of phonons on 2 dimensional ($2D$)  plane, we take the state vector in quaternion projected space. A quaternion $h\in {\bf H}$ is described as
\begin{eqnarray}
h&=&\alpha{\bf I}+\beta{\bf i}+\gamma{\bf j}+\delta{\bf k}\nonumber\\
&=&\alpha\sigma_0+\beta \sqrt{-1}\sigma_1+\gamma \sqrt{-1}\sigma_2+\delta\sqrt{-1}\sigma_3\nonumber\\
&=&\left(\begin{array}{cc}
    \alpha+\sqrt{-1} \delta& \gamma+\sqrt{-1}\beta\\
    -\gamma+\sqrt{-1} \beta& \alpha-\sqrt{-1}\delta\end{array}\right),
\end{eqnarray}
where $\sigma_i$ $(i=1,2,3)$ are the Pauli matrices, and\\
${\bf i}^2={\bf j}^2={\bf k}^2=-1$, ${\bf i} {\bf j}=-{\bf j}{\bf i={\bf k}}$, ${\bf j} {\bf k}=-{\bf k}{\bf j}={\bf i}$, ${\bf k} {\bf i}=-{\bf i}{\bf k}={\bf j}$.

Using complex coordinates $z, w\in{\bf C}$, one can write
\begin{equation}
{\bf H}=\{\left(\begin{array}{cc}
                              z & w\\
                              -\bar w&\bar z\end{array}\right)\}.
\end{equation}
Quaternions $p_1$ and $p_2$ are said to be equivalent if there exists $h\in{\bf H}\setminus\{0_{\bf H}\}$ such that $h p_1=p_2 h$. We consider propagation of solitons in a $2D$ plane.

Structure of this presentation is as follows. In sect..2, we summarize the principle of TR-NEWS: time-reversal based nonlinear elastic wave spectroscopy. In sect. 3, we explain setup of transducers and receivers.
In sect.4 convolution of the KhZa wave function and its TR wave function is explained.
 Quaternion neural network and its topological properties are explained in sect. 5. Details of Altland-Zirnbauer $DIII$ class in $(2+1)D$ spacetime is given in sect.6. 
In sect.7,  we present mathematical bases of Quaternion Fourier Transforms (QFT), using the fact that quaternions are Clifford numbers\cite{Chevalley46}.
Discussion and perspective are given in sect. 8.

\section{Symmetries in propagation of solitary waves in matters}
\label{sec:1}
Time reversal symmetry based nonlinear elastic wave spectroscopy, in which one optimizes the convolution of the scattered wave from defects in materials and its time reversed wave show peaks was an effective method for non-destructive testing (NDT)\cite{BLEFF16}.

Goursolle et al.\cite{GCDSBM07} studied propagation of nonlinear elastic waves in materials with hysteresis. The hysteretic nonlinearity model was based on Preisach-Mayergoyz space (PM space)\cite{Mayergoyz86}. 
In $2D-$Kelvin notation model, $3D-$Kelvin stress vector is defined as
\begin{equation}
\tilde \tau=\left(\begin{array}{c}
                      \tau_{xx}\\
                      \tau_{yy}\\
                      \tau_{xy}\sqrt 2\end{array}\right)=\left(\begin{array}{ccc}
                                                                  C_{11}& C_{13} & 0\\
                                                                  C_{13}& C_{33} & 0\\
                                                                  0 & 0 & C_{44} \end{array}\right)\left(\begin{array}{c}
                                                                                              \epsilon_{xx}\\
                                                                                              \epsilon_{zz}\\
                                                                                              \epsilon_{xz}\end{array}\right)
\end{equation}
where Hooke's law $\tau_{ij}=C_{ijkl}\epsilon_{kl}$, $c_{ijkl}$ is the elastic coefficients, and 
\begin{equation}
\epsilon_{kl}=\frac{1}{2}(\frac{\partial u_k}{\partial x_l}+\frac{\partial u_l}{\partial x_k})
\end{equation}
is the strain tensor.

The Newton's second law is
\begin{equation}
\frac{\partial v_i}{\partial t}=\frac{1}{\rho_0}\frac{\partial\tau_{ij}}{\partial x_j}.
\end{equation}
Phenomenologically stress on a $2D$ plane have components parallel to the axis $\tau_{xx}, \tau_{yy}$ and have nonzero angles to the two axes $\tau_{xy}$, strain tensors $\epsilon_{kl}$ have components with both legs tangential to the 2D plane, both legs orthogonal to the plane and one leg tangential and one leg orthogonal.
Distance squared between two points $dl^2={dx^2+dy^2+dz^2}$ before stress changes to $dl'^2=dl^2+2\epsilon_{ik}dx_i dx_k$\cite{LL64},
\begin{equation}
\epsilon_{ik}=\frac{1}{2}(\frac{\partial u_i}{\partial x_k}+\frac{\partial u_k}{\partial x_i}+\frac{\partial u_l}{\partial x_i}\frac{\partial u_l}{\partial x_k}).
\end{equation}

For nonlinear acoustic wave spectroscopy in media with hysteresis, one can use memristor that creates forward propagating and backward propagating waves, and measure scattered waves\cite{BFMT18}. When transducers and receivers are displayed on a $2D$ plane, quaternion neural network may help optimization of NDT\cite{DSF16,DSM18,DSFN18,SF20}.
In propagation of acoustic waves in non linearly oscilating media, solitary wave property appears\cite{SSB17}. In the analysis of TR-NEWS  in NDT, a proper choice of strengths and intervals of impulse from transducers may allow detection of gravitational effects.

 In generalized pulse inversion methods of TR-NEWS \cite{DSP08,VPDS09,DSVP09,DSP20}, excitation function bases are taken as 
\begin{equation}
x_E=x(t), x_\epsilon=x(t)e^{i 2\pi/3}, x_{\epsilon*}=x(t)e^{-i 2\pi/3},
\end{equation}
and their response are $y_E, y_{\epsilon}, y_{\epsilon*}$ respectively. 

In neural networks, function on imput layers are
\begin{equation}
x_A(t)=-\frac{1}{2}x(t), x_{B1}(t)=\frac{\sqrt{3}}{2}x(t), x_{B2}(t)=-\frac{\sqrt{3}}{2}x(t),
\end{equation}
and that on output layers are $y_A$, $y_{B1}$, $y_{B2}$, respectively.

 Nonlinear responses are parametrized as
\begin{equation}
y(t)=N\,L[x(t)]=N_1 x(t)+N_2 x^2(t)+N_3 x^3(t).
\end{equation}
The nonlinear responses on output layers are expressed as
\begin{eqnarray}
N_3 x^3(t)&=&\frac{4}{3}[y_E (t)+2 y_A (t)-y_{B1}(t)-y_{B2}(t)]=s_3(t),\nonumber\\
N_2 x^2(t)&=&\frac{2}{3}[y_{B1}(t)+y_{B2}(t)]=s_2(t),\nonumber\\
N_1 x(t)&=&y_E(t)-s_2(t)-s_3(t)=s_1(t).
\end{eqnarray}

To obtain the energy flow to the receiver from $y_E, y_A, y_{B1}$ and $y_{B2}$, 
\begin{eqnarray}
&&N_1^2\int_{-\infty}^\infty|x(t)|^2 dt,\nonumber\\
&&N_2^2\int_{-\infty}^\infty|x^2(t)|^2 dt,\nonumber\\
&&N_3^2\int_{-\infty}^\infty|x^3(t)|^2 dt,
\end{eqnarray} 
there is DORT(D\'ecomposition de l'Op\'erateur Retournement T\'emporell) method \cite{VPDS09, DSVP09}.

 The Fourier transform ${\mathcal F}s_1(\omega)$ has a peak around several hundred kHz dependent on chirp coded excitation 
\begin{equation}
c(t)=A\cdot \sin(\psi(t))
\end{equation}
where $\psi(t)$ is linearly changing instantenious phase of the order of a few MHz. Analyses of nonlinear solitary wave in complex media are performed in\cite{SMJCBDS14}.

 When the impulse resonse of the medium is expressed by $h(t-t',T)$ where $T$ is the time duration, the response is expressed by convolution
\begin{equation}
y(t,T)=h(t)*c(t)=\int_{\bf R}h(t-t',T)c(t')dt'.
\end{equation}
The response of the TR wave is given by the convolution
\begin{equation}
y_{TR}(t , T)=\int_{\Delta t}y(T-t'-t)c(T-t')dt'*h(t)\sim \delta(t-T),
\end{equation}
which is a linear combination of $s_1(t),s_2(t),s_3(t)$.

 In the calculation of convolutions we use semigroup splitting method of Trotter \cite{Trotter58}. The differential equation $\dot{y}=f(y)$ in ${\bf R}^n$ is split as $\dot{y}=f^{[1]}(y)+f^{[2]}(y)$, and exact flows $\varphi_t^{[1]}$ and $\varphi_t^{[2]}$ of $\dot{y}=f^{[1]}(y)$ and $\dot{y}=f^{[2]}(y)$, respectively are calculated. 
The functions
\begin{equation}
\mit\Phi_h^*=\varphi_h^{[2]}\circ \varphi_h^{[1]},\quad \mit\Phi_h=\varphi_h^{[1]}\circ \varphi_h^{[2]},
\end{equation}
connects initial value $y_0$ and final value $y_2$, via different paths.
\begin{figure}
\begin{center}
\includegraphics[width=4cm,angle=0,clip]{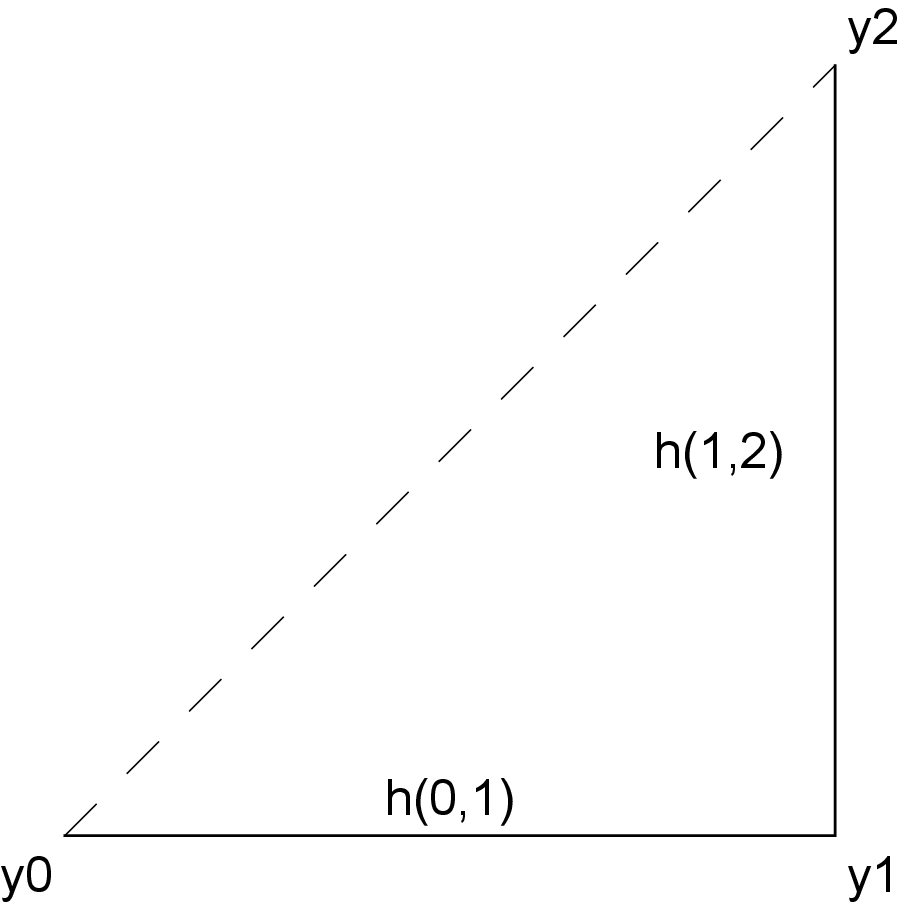}
\caption{Path dependence of outputs. }
\label{fig:1}       
\end{center}
\end{figure}
%
\begin{figure}
\begin{center}
\includegraphics[width=4cm,angle=0,clip]{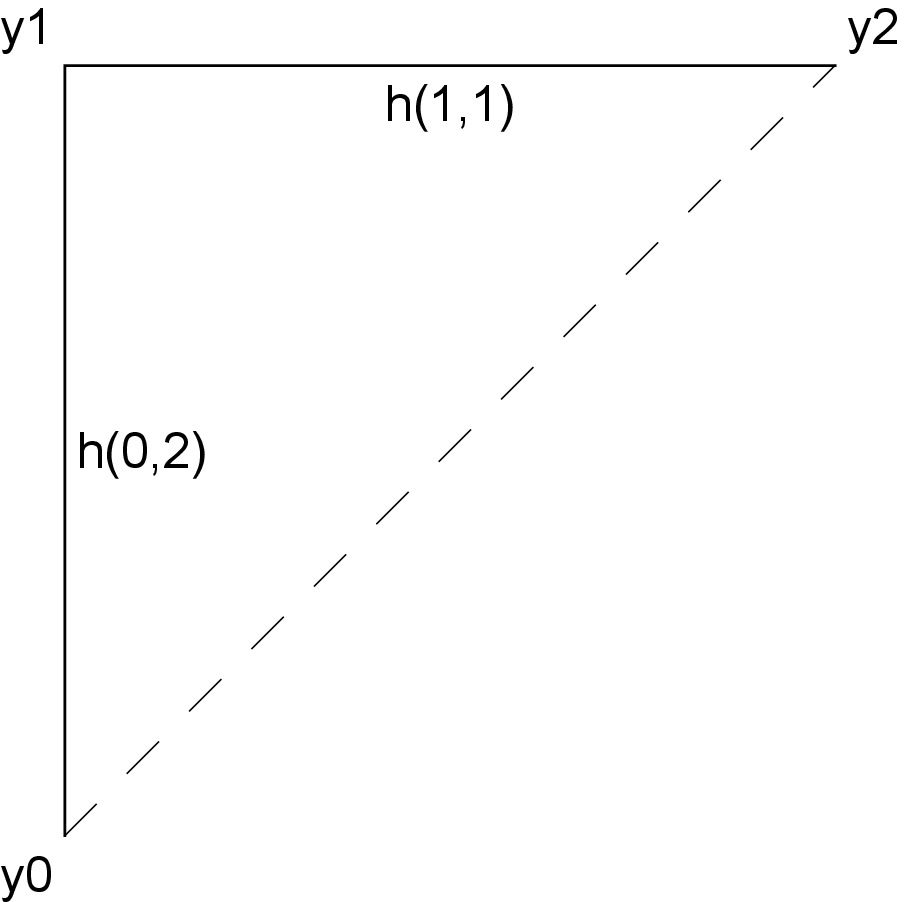} 
\caption{Path dependence of outputs.}
\label{fig:2}       
\end{center}
\end{figure}
The formula $e^{A+B}=\displaystyle\lim_{n\to\infty}(e^{A/n} e^{B/n})^n$ does not follow if $A,B$ are non-commutative. However Trotter\cite{Trotter58} showed that $T_t f(x)=f(x-t)$ and $T_t' f(x)=f(x+t)$ form semi-groups, and due to the Hille-Yosida's theorem\cite{Yosida48,Yosida58,HP58,YKI67} 
\begin{equation}
S_t f(x)=\lim_{h\to 0}(T_h T_h' )^{[t/h]} f(x).
\end{equation}
can be defined in discrete time steps.

If the semi-group of linear operators $T_t$ satisfy
\begin{eqnarray}
&1)&  T_0=I, T_t T_s=T_{t+s},\nonumber\\ 
&2)&  s\cdot\lim T_t x=T_{t_0} x, (x\in X),\nonumber\\
&3)&  \parallel T_t\parallel\leq e^{\beta|t|} (\beta>0),
\end{eqnarray}
where $s\cdot\lim$ means strong convergence limit, then operators $\{T_t\}$ form a group in Banach space $X$.

Hille-Yosida theorem\cite{Yosida48,Yosida58,HP58,YKI67} says that for finite opertor $T_t'$ for $t>0$, 
\begin{equation}
\overline{\lim_{|\nu|\to\infty} }\parallel ((\lambda+\sqrt{-1}\nu) I -A)^{-1}\parallel <\infty,
\end{equation}
if $\lambda>\beta$ is satisfied the operator $T_t'$ can be extended to a regular operator $T_{t+\sqrt{-1}s}$ in the cone area
\begin{equation}
\{ t+\sqrt{-1}s; |s|\leq ct, c, {\rm is, real, postive}\}.
\end{equation}

Operators $\{T_t\}$ form a semi-group with infiniesimal generator $A$
\begin{equation}
s\cdot\lim_{h\to 0}h^{-1}(T_h-I)x=A x.
\end{equation}

The $i$th layer has connection to $j$th hidden layer 
\begin{equation}
h(i,j)=h(i-1,1)\cdot h(i-1,2)\quad \forall j\in\{1,2\}.
\end{equation}

 If the semi-group of linear operators $T_t$ satisfy
\begin{eqnarray}
&1)&  T_0=I, \quad T_t T_s=T_{t+s},\nonumber\\ 
&2)&  s\cdot\lim T_t x=T_{t_0} x, (x\in X),\nonumber\\
&3)&  \parallel T_t\parallel\leq e^{\beta|t|} (\beta>0)
\end{eqnarray}
where $s\cdot\lim$ means strong convergence limit, then operators $\{T_t\}$ form a group in Banach space $X$.

\section{Setup of memosducer\\ ( transducers) and receivers}
\label{sect:2}
Let us consider memosducers or transducers which have hysteresis and emits sonic beams and TR sonic beams.  Memosducers were proposed in\cite{DSFN18} for a use of TR signal processing.
The fact that hysteresis occurs from non holonomicity was explained in \cite{SF20a}.

Parallel transformations have a special meaning in Clifford algebra. Lounesto \cite{Lounesto93,Lounesto01} defined the basis\\
 $\{1,e_1, e_2, e_1\wedge e_2\}$ for $\wedge V$, that satisfy
the following multiplication table
\begin{center}
\begin{tabular}{|c|c|c|c|}
\hline
$ \wedge$ & $e_1$ & $e_2$ & $e_1 \wedge e_2$ \\
\hline
$ e_1$ & 0 & $e_1\wedge e_2$ & 0\\
 $e_2$ &  $ -e_1\wedge e_2$ & 0 & 0\\
 $e_1\wedge e_2$& 0 & 0 & 0 \\
 \hline
 \end{tabular}
\end{center}
and the second product $\dot \wedge$ whose multiplication table is the following.
\begin{center}
\begin{tabular}{|c|c|c|c|}
\hline
$ \dot \wedge$ & $e_1$ & $e_2$ & $e_1\wedge e_2$\\
\hline
$e_1$&0 & $e_1\wedge e_2+b$ &  $-b e_1$\\
$e_2$&$- e_1\wedge e_2-b $&0&$-b e_2$\\
$e_1\wedge e_2$& $-b e_1$  &$-b e_2$ & $-b^2-2b e_1\wedge e_2$ \\
 \hline
 \end{tabular}
 \end{center}
 where $b>0$ characterizes parallel shifts. The multiplication table can be rearranged to the following.
 \begin{center}
\begin{tabular}{|c|c|c|c|}
\hline
$ \dot \wedge$ & $e_1$ & $e_2$ & $e_1\wedge e_2 +b$\\
\hline
 $e_1$&0 & $e_1\wedge e_2+b$ &   0\\
$e_2$&$ -e_1\wedge e_2-b$ &0& 0\\
$e_1\wedge e_2+b$& 0  &0  & 0 \\
 \hline
 \end{tabular}
 \end{center}
 It means that in $(2+1)D$, shifts in $e_1\wedge e_2$ can be treated by a simple coordinate transformation.
 
Effects that appear by identifying parallel lines along the horizontal axis from $A$ and $B$ are regarded as instantons, although the presence of the global symmetry is not evident.

We arrange memosducers on the left wall of the $2D$ material equally spaced and receivers on the right wall equally separated. By adjusting memoducers, solitonic wave from a transducer $T_i$ and time reversed (TR) solitonic wave propagate on a $2D$ plane $(\hat u,\hat v)$. At $t=0$, the wave front is at $x=0$, and propagate within the cone in $t>0$ region. The TR wave propagate within the cone in $t<0$ region. We assume time reversal invariance in the recurrent steps.

\begin{figure} [htb]
\begin{center}
\includegraphics[width=6cm,angle=0,clip]{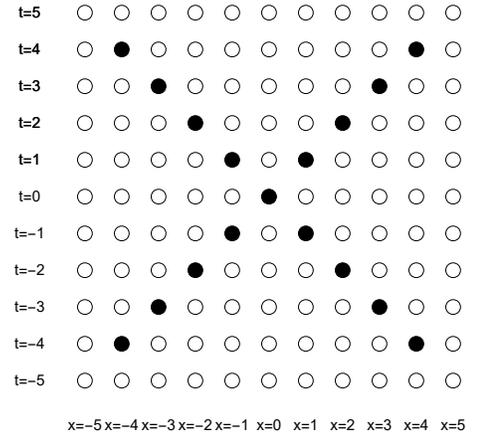} 
\label{fig:3}
\caption{Wave front of solitonic waves. }
\end{center}
\end{figure}

We consider phonons produced at $x=0$ at time $t=0$ and propagate forward and backward with a scaled velocity, as shown in Fig.4. In order to reduce effects of boundary conditions we add padding layers at $t=\pm 5$ and $x=\pm 5$. The number of padding layers is to be changed according to accumulated data.

Using the notation of \cite{Aggarwal18}, we choose for transducers on a line and receivers on a line $L_1=5, B_1=1$, and for forward and backward propagation $d_1=2$, and the filter $F_1=2$. $L_2=L_1-F_1+d_1=4$,
$B_2=B_1-F_1+d_1=0$. The filter has the size $2\times 1\times 2$.

As shown in Fig. 3, nonlinear sound waves and their TR waves emitted from transducers $T_1,\cdots,T_N$ on the left walls are received by receivers $R_1,\cdots,R_N$ on the right wall. Sound waves are scattered by an object shown by a black disk between the walls and on a plane on which transducers and receivers are placed. Dashed lines between transducers and receivers have longer paths than solid lines, whose information is contained in larger time delay $\tau_i$ in signals of receivers $R_i$. 

The waves from $T_i$ are scattered by cracks in the medium if they exist, and solitonic waves are disturbed, 
and they are received by receiver $R_j$. We consider the situation of 5 transducers and 5 receivers
\begin{figure}
\begin{center}
\includegraphics[width=6cm,angle=0,clip]{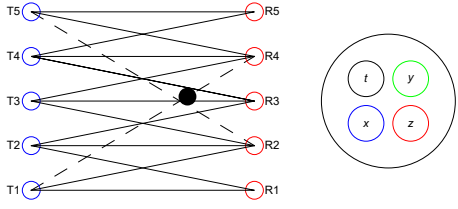}%
\label{fig:4}
\end{center}
\caption{Networks of 5 transducers $T_1,\cdots,T_5$ and 5 receivers $R_1,\cdots, R_5$. Trajectories are expressed by quaternions ${\bf H}=\tau{\bf I}+x{\bf i}+y{\bf j}$, where $\tau,x,y$ are real and can be mapped to $M(2, {\bf C})$.}
\end{figure}
Taking the length $x$ in the unit of $c_0 t$, the wave front which was at $x=0$ at $t=0$ propagates to $x\geq 1$, $t\geq 1$. For TR waves the wave front propagates to $x\geq 1$, $t\leq -1$. 
\begin{figure}
\begin{center}
\includegraphics[width=6cm,angle=0,clip]{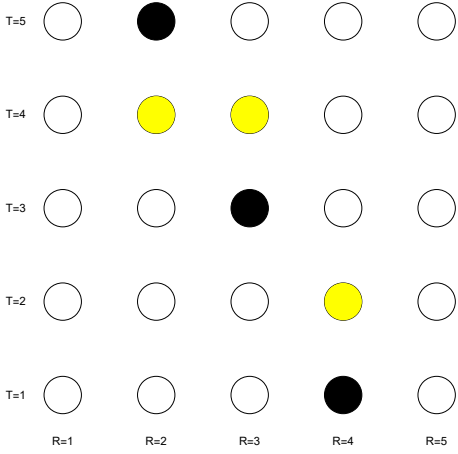}%
\label{fig:5}
\end{center}
\caption{Disturbance of beams between transducers $T_i$ and receivers $R_j$ on the $2D$ plane shown in Fig.4. Black disks correspond to strong disturbances and yellow disks correspond to weak disturbances. }
\end{figure}

In the $2D$ plane on which transducers and receivers are placed $h=\tau{\bf I}+x{\bf i}+y{\bf j}$, $q_1=(t+l_{mm}/c){\bf I}+X{\bf i}+Y{\bf j}$, $q_2=(t-l_{mm}/c){\bf I}+X{\bf i}+Y{\bf j}$.  
\begin{eqnarray}
&&h q_1-q_2 h=\nonumber\\
&&\left(\begin{array}{cc}
2\frac{l_{mm}}{c}(\tau+\sqrt{-1}x) &- 2(\frac{l_{mm}}{c} y +\sqrt{-1} Z)\\
-2(\frac{l_{mm}}{c} y-\sqrt{-1}Z) & 2\frac{l_{mm}}{c}(\tau-\sqrt{-1}x)
\end{array}\right)
\end{eqnarray}
\label{eq2D}
where $Z=Xy-xY$.

A real trajectory outputs $y_j$ has the local partial derivative $z(m,n)=\frac{\partial y_m}{\partial y_n}$,
whose products gives variation of real outputs $o$ with respect to weight functions: 
\begin{equation}
\frac{\partial o}{\partial w}=\sum_{P\in{\mathcal P}}\prod_{(m,n)\in {\mathcal P}} z(m,n).
\end{equation}
Here the $\mathcal P$ is the aggregates of paths containing $\partial y_m$ and $\partial y_n$.

Relativistic dynamics of the Maxwell-Einstein equation can be represented by instant form, front form and point form\cite{Dirac49}. The usual parametrization $t$ in instant form is defined by parallel transformation from the system that satisfy $p^\sigma p_\sigma-{\mathcal M}^2=0$ at the Lorentz coordinate $u_0=0$, while parametrization $\tau=t-x/c_0$ in KhZa dynamics is front form. The propagation of massless partcles in the front form is shown in Fig.6 and the propaggation of massive particle and massless particle in the point form is shown in Fig.7.

Corresponding to the presence of hysteresis, in $(2+1)D$ space we take the position of transducer $T_m$ and receiver $R_m$ as in Fig. 8 and the propagation direction dependence of massive phonon wave fronts are shown in Fig.9.

When a $(2+1)D$ front form wave function is expressed by quaternions $q\ne 0$, equivalent quaternions that satisfy $q_1 q=q q_2$ have the periodicity in $\tau$ direction by $2 l_{mn}/c$.  When phonons have effective mass due to scattering in media, the wave front in instatnt form changes to the point form. The point form comes from taking a branch of hyperboloid $u^\rho u_\rho=\kappa^2$, $u_0>0$.
\begin{figure}[htb]
\begin{center}
\includegraphics[width=6cm,angle=0,clip]{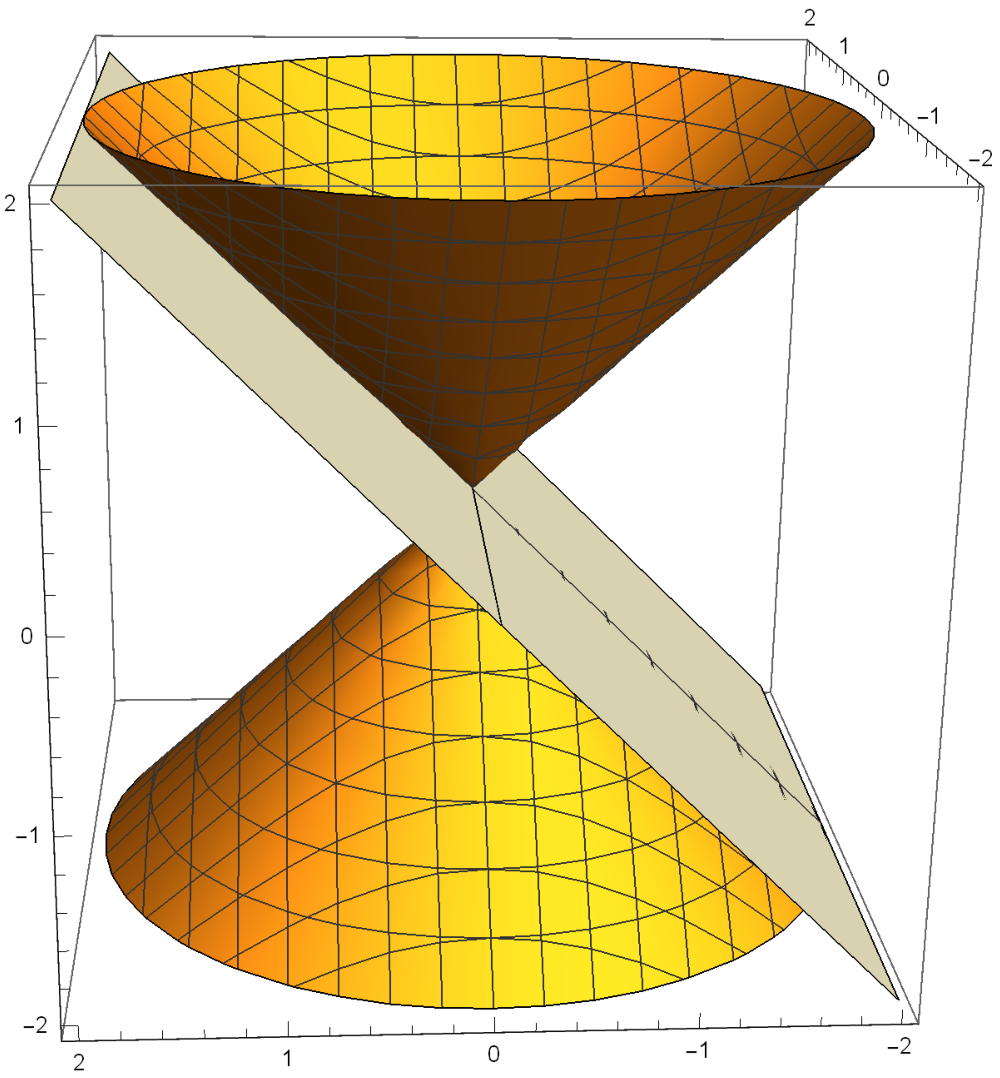}
\label{fig:6}
\caption{The $(2+1)D$ wave front of massless particles in the front form. The vertical line is the time axis and the plane is $t\pm l_{mn}/c=0$ frame.}
\includegraphics[width=6cm,angle=0,clip]{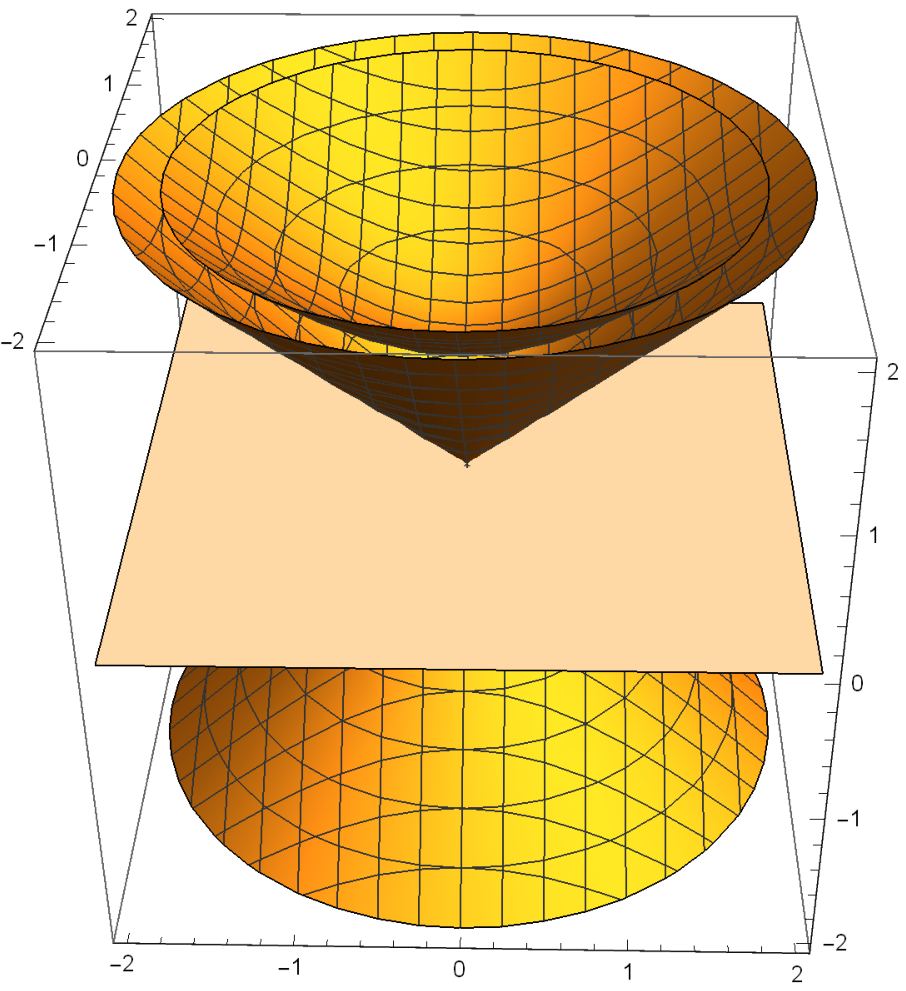}
\label{fig:7}
\caption{The $(2+1)D$ wave front of massive particle in the instant form, which exists inside the cone of massless particles. The plane is the frame of $t=0$}
\end{center}
\end{figure}

 Actions depending on paths yields hysteresis which is related to transformation of holonomy groups under parallel displacements.   
\begin{figure} [htb]
\begin{center}
\includegraphics[width=6cm,angle=0,clip]{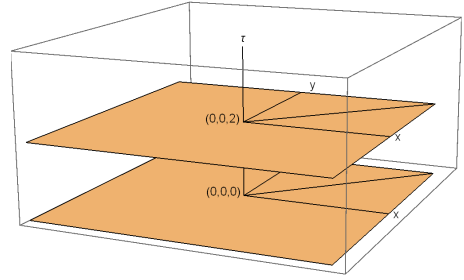} 
\label{fig:8}
\end{center}

\caption{The $2D$ wave propagation direction dependence on the transducer position $T_m$ at $(0,0,0), (0,0,2)$ and receiver position $R_n$: $(m,n)$.}
\end{figure}
\begin{figure}
\begin{center}
\includegraphics[width=6cm,angle=0,clip]{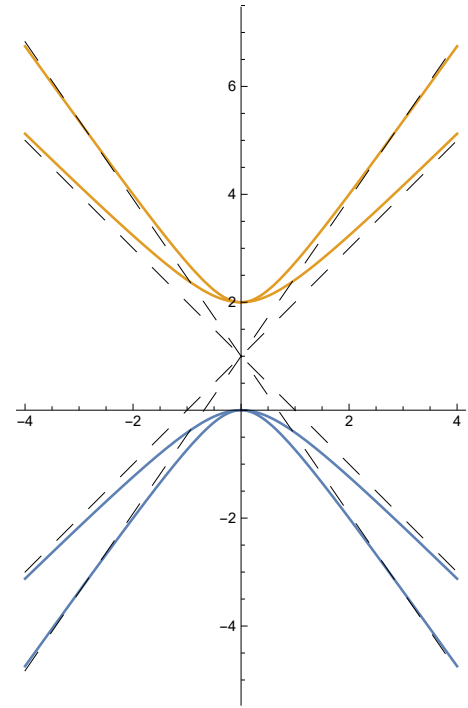}%
\label{fig:9} 
\end{center}
\caption{
The propagation direction dependence of point form wave fronts with non-zero effective mass. Vertical ordinate is $\tau$, abscissa is the distance from the origin. }
\end{figure}

We use instead of $t$, a step parameter $\tau=t\pm l_{m,n}/c$, where $l_{mj}$ is the length of the path between the transducer $T_m$ and receiver  $R_n$ on the $(X, Y)$ plane.

 In order to get positions of scatterers, we define the Loss function for input $X$, output $Y$ 
\begin{equation}
L=\sum_{(X,Y)\in D}(Y-\hat Y)^2+\lambda\sum_{i=0}^d w_i^2,
\end{equation}
where $d=3$ is the degree of bases function, $\hat Y=\sum_{i=0}^d w_i X^i$, and $\lambda$ is a constant.
Following Aggarwal \cite{Aggarwal18}, we denote $(w_1,\cdots,w_d)=\bar W$.

 Consider $g_1(\cdot), g_2(\cdot),\cdots,g_N(\cdot)$ computed in layer $m$, and composition function in layer $m+1$ is $f(g_1(\cdot),\cdots, g_N(\cdot))$, and for an input $X$, weight $w$, consider input $f(w)$ and split paths from an input layer to an output layer $Y$ and hidden layer $Z$ by 
\begin{equation}
Y=f(w), p=g(Y) \quad {\rm and \quad} Z=f(w), q=h(Z).
\end{equation}

In our application $X$ and $Y$ are descibed by a KhZa soliton wave functions propagating in the definite direction.

 The partial derivative of output with respect to $w$ is
\begin{eqnarray}
&&\frac{\partial o}{\partial w}=\frac{\partial o}{\partial p}\cdot \frac{\partial p}{\partial w}+\frac{\partial o}{\partial q}\cdot \frac{\partial q}{\partial w},\nonumber\\
&&=\frac{\partial o}{\partial p}\cdot \frac{\partial p}{\partial y}\cdot \frac{\partial y}{\partial w}+\frac{\partial o}{\partial q}\cdot \frac{\partial q}{\partial z}\cdot \frac{\partial z}{\partial w},\nonumber\\
&&=\frac{\partial K(p,q)}{\partial p}\cdot g'(y)\cdot f'(w)+\frac{\partial K(p,q)}{\partial q}\cdot h'(z)\cdot f'(w).\nonumber\\
\end{eqnarray}

We define $w(X_m,Y_n)$ defined by the path from input position $X_m$ of transducer to the output position $Y_n$ of the receiver. When there is a scatterer in the $2D$ plane as shown in the figure, among functions $w(X_m,Y_n)$, $w(3,3)$ and $w(4,3)$ are expected to have large disturbances. 
With wider range of filters of $|X_m-X_n|\leq 3$ $w(1,4)$ and $w(5,2)$ will have large disturbances.

 Since there are $X_E, x_{\epsilon}$ and $x_{\epsilon^*}$ bases we take $p=3$.

One of the aim of this research is to find optimal functions $f(w), g(y), h(z)$ and $z(i,j)$, such that the loss function $L$ becomes small. 

 We define outputs $Y_n$ in the forward phase, using hidden layer variables $h(i,q)$, where $q=1,3$ and $i$ defines the recurrence order.  The hidden layer variable $h(i,q)$ and $h^{TR}(i,q)$ distinguish interference with original or TR phonon beams. Relation between outputs and hidden layer variables are
\begin{eqnarray}
h(i,q)&=&\alpha(W_{hh}\otimes h(i-1,q)+W_{hx}\otimes X_i+b_h),\nonumber\\
\alpha(Q)&=&f(r){\bf I}+f(x){\bf i}+f(y) {\bf j},\nonumber\\ 
Y_i&=&\beta(W_{hy}\otimes h(i,q)) ,
\end{eqnarray}
where $\alpha$ and $\beta$ are split activation functions. $b_h$ is the bias of the hidden state.

For calculation of $\bar Y=[y_1,\cdots,y_n]^T$, we choose training sets $\hat Y=[\hat y_1,\cdots,\hat y_n]$,
that satisfy $\hat y_i=\bar H_i \bar W^T=\sum_{j=1}^n w_j\Phi(\bar X_i)$. Here $\bar H_i$ is $m$ dimensional and represents in the hidden layer, yields $\hat y$ for the $i$-th training point $\bar X_i$.

The hidden layer weight function for original waves $\bar h(i,q)$ and for TR waves can be taken as
\begin{eqnarray}
\bar h(i,q)&=&\tanh(W_{xh}\bar X_i+W_{hh}\bar h(i-1,q)),\nonumber\\
\bar h^{TR}(i,q)&=&\tanh(W_{xh}^{TR}\bar X_i+W_{hh}^{TR}\bar h^{TR}(i+1,q)), \nonumber\\
\bar Y(i,q)&=&W_{hy}\bar h(i,q)+W_{hy}^{TR}\bar h^{TR}(i+1,q),\nonumber\\
\hat Y_i&=&\bar W\cdot\bar X_i.
\end{eqnarray}

The partial derivatives of the loss function are given by the trained hidden layer function $h(i,q)$.
\begin{eqnarray}
&&\frac{\partial L_i}{\partial W_{hy}^r}=\frac{\partial L_i}{\partial y_i^r} \frac{\partial y_i^r}{\partial W_{hy}^r}
+\frac{\partial L_i}{\partial p_i^{\bf i}} \frac{\partial p_i^{\bf i}}{\partial W_{hy}^r}
+\frac{\partial L_i}{\partial p_i^{\bf j}} \frac{\partial p_i^{\bf j}}{\partial W_{hy}^r}\nonumber\\
&&=(p_i^r-y_i^r) h^r(i,q)+(p_i^{\bf i}-y_i^{\bf i}) h^{\bf i}(i,q)+(p_i^{\bf j}-y_i^{\bf j}) h^{\bf j}(i,q),\nonumber\\
&&\frac{\partial L_i}{\partial W_{hy}^{\bf i}}=(p_i^r-y_i^r) (-h^{\bf i}(i,q))+(p_i^{\bf i}-y_i^{\bf i}) h^r(i,q),\nonumber\\
&&\frac{\partial L_i}{\partial W_{hy}^{\bf j}}=(p_i^r-y_i^r) (- h^{\bf j}(i,q))+(p_i^{\bf j}-y_i^{\bf j}) h^r(i,q).
\end{eqnarray}

 Here $p_i=\beta(W_{hy}\otimes h(i,q))$ is calculated by split activation functions and  hidden weight matrix for $K$ time steps,
\begin{equation}
\frac{\partial L}{\partial W_{hh}}=\sum_{i=1}^K \frac{\partial L_i}{\partial W_{hh}}.
\end{equation}

At each $i$th time step,
\begin{equation}
\frac{\partial L_i}{\partial W_{hh}}=\sum_{m=1}^i (\frac{\partial L_m}{\partial W_{hh}^r}+
\frac{\partial L_m}{\partial W_{hh}^{\bf i}}{\bf i}+\frac{\partial L_m}{\partial W_{hh}^{\bf j}}{\bf j}).
\end{equation}

Input weight matrix is $\displaystyle\frac{\partial L}{\partial W_{hx}}=\displaystyle{\sum_{i=1}^K }\frac{\partial L_i}{\partial W_{hx}}$.

Hidden biases is $\displaystyle\frac{\partial L}{\partial b_{h}}=\displaystyle{\sum_{i=1}^K} \frac{\partial L_i}{\partial b_{h}}$.

The update of $\bar W$ can be written as $\bar W\Leftarrow \bar W(1-\alpha\cdot \lambda)+\alpha(Y_i-\hat Y_i)\bar X$.

Mapping of metric data $\varphi (x_1,x_2,t)\to \varphi(x_1,x_2,N_1,N_2,\tau)$ ( transformation from the instant form to the front form)  allows  decomposition of wave function as $\varphi=\varphi^++\varphi^-$, where
\begin{equation}
\varphi^+=\frac{1}{2}(\varphi+\sqrt{-1} N_\perp\cdot\varphi), \quad \varphi^-=\frac{1}{2}(\varphi-\sqrt{-1} N_\perp\cdot\varphi)
\end{equation}
where $N_\perp$ is orthnormal to the wave front defined on the $(x_1,x_2)$ plane.

We calculate propagation of a pulse $f(x_1,x_2)$ and the TR pulse $\hat f(x_1,x_2)$, expressed by inputs $X,\hat X$ of $T_m$, and expressed by outputs $Y,\hat Y$ of $R_n$. 

In the case of propagation of phonons in hysteretic media, the output at time $t$ is given by extending the double integral to that in Preisach-Mayergoyz space \cite{Mayergoyz86,BT10}.

\section{Convolution of the KhZa wave function and its TR wave function }
\label{sec:3}
In this section we explain the mechanism of TR-NEWS using the soliton wave function of Lapidus and Rudenko\cite{LR84,LR92}. 

They showed a spectral decomposition of the fuction $V$ of eq. (1-3) as
\begin{eqnarray}
V&=&\sum_{n=1}^\infty\frac{2}{n z}J_n[\frac{n z}{\sqrt{1+N^2 z^2}} exp(-\frac{R^2}{1+N^2 z^2})] \nonumber\\
&&\times \sin[n(\theta+\tan^{-1}(N z)-\frac{R^2 N z}{1+N^2 z^2})],
\end{eqnarray}
where $R^2$ is a constant defined by $N$ which characterises nonlinearity of the material, and the shock formation coordinate $z_s$.

As a test, we take $n=1,2,3$ and $N=0.5, 0\leq z\leq 20$, $R^2=1$ and $c_0=1$. 
For smaller $R^2$, $z_s$ becomes smaller.

If there are singularities in finite dimensional wavefuncions, the convolution of distributions\cite{Schwartz61,Hoermander83a} 
is a useful tool. Since we know analytical solutions of KhZa nonlinear differential equation, we first consider $(1+1)D$ convolution, before $(2+1)D$ convolutions of real functions.
\subsection{Convolution of $(1+1)D$ real wave functions}
\label{subsec:1}
Numerical solution of fluid dynamics in $(1+1)D$ space was studied by Khelil et al.\cite{KMPP01}.
 The variable $t=0$ yields $\tau=t-z$, $\theta=\omega\tau=\omega t-\omega z$. 
 We consider cases of $\theta=\omega \tau=0, \pi/10, 3\pi/10, 5\pi/10$. Parameters $\omega$ and $z$ are dependent on phonetic beams of transducers and directions of the beams relative to receivers.

 \begin{figure}[htb]
\begin{center}
\includegraphics[width=6cm,angle=0,clip]{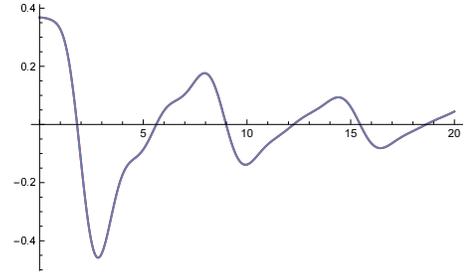}
\end{center}
\caption{$V_{123}^-$ wave function of $\tau=t-z$ (red)  and $V_{123}^+$ wave function of $\tau=t+z$ (blue)  for $\omega\tau=\pi/2$.  The abscissa is $z$.}
\label{fig:10}
\end{figure}
\begin{figure}[htb]
\begin{center}
\includegraphics[width=6cm,angle=0,clip]{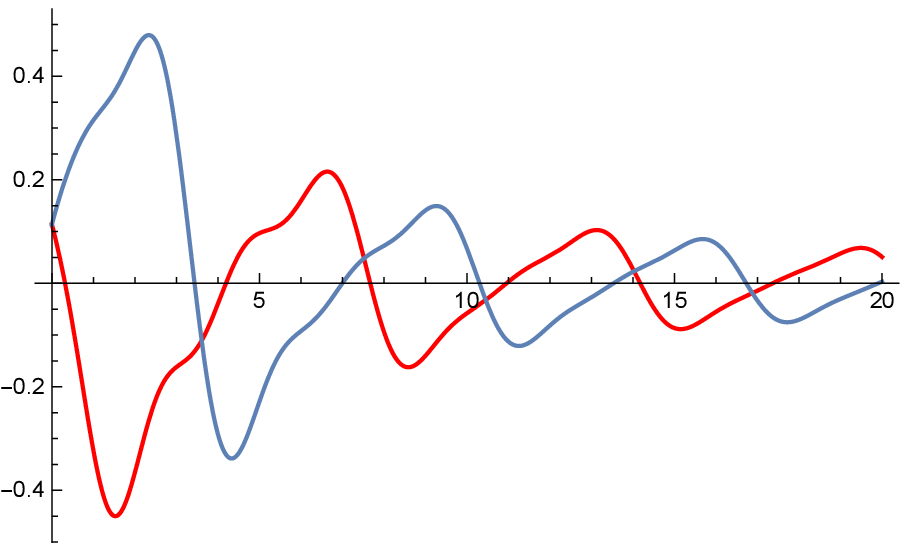} 
\end{center}
\caption{$V_{123}^-$ wave function of $\tau=t-z$ (red) and $V_{123}^+$ wave function of $\tau=t+z$ (blue) for $\omega\tau=\pi/10$. The abscissa is $z$. }
\label{fig:11}
\end{figure}
\begin{figure}[htb]
\begin{center}
\includegraphics[width=6cm,angle=0,clip]{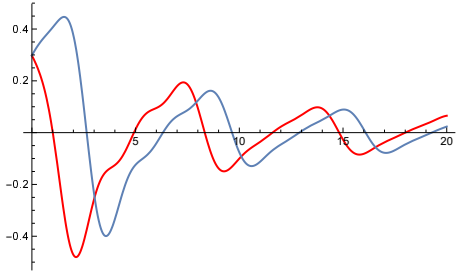}
\end{center}
\caption{$V_{123}^-$ wave function of $\tau=t-z$ (red) and $V_{123}^+$ wave function of $\tau=t+z$ (blue) for $\omega \tau=3\pi/10$. The abscissa is $z$. }
\label{fig:12} 
\end{figure}

When $\omega\tau=0$ there appears hysteretic effects between $V^-_{123}$ defined by $\tau=t-z$ and $V^+_{123}$ defined by $\tau=t+z$, but when $\omega\tau=\pi/2$, $V^-_{123}$ and $V^+_{123}$ are identical.

The Fourier transform of $V_{123}^-$ contains lower and broader peak spectra than those in the convolution of $V_{123}^-$ and $V_{123}^+$. The convolution of $V^+_{123}$ and $V^-_{123}$ for $\omega\tau=0$ and $\pi/2$ are almost same, but for $\omega\tau=\pi/10$ and $3\pi/10$, the height of peakes are almost the same, but they the shape of sidelobes are different. The convolution was calculated for $z\in [1,256]$ by using a library program of a supercomputer at RCNP.
 Discrete Fourier transformations were done by using Mathematica\cite{Mathematica12}. 
\begin{figure}[htb]
\begin{center}
\includegraphics[width=6cm,angle=0,clip]{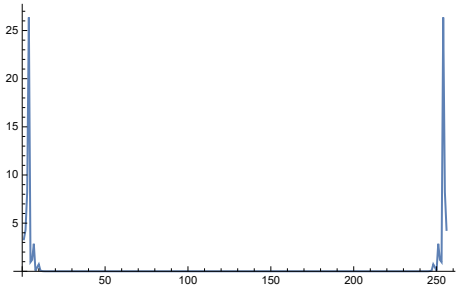} 
\end{center}
\caption{Convolution of $V_{123}^+$ and $V_{123}^-$ for $\omega\tau=0$. Absolute values of the numerical Fourier transform are plotted. }
\label{fig:13}
\end{figure}
\begin{figure}[htb]
\begin{center}
\includegraphics[width=6cm,angle=0,clip]{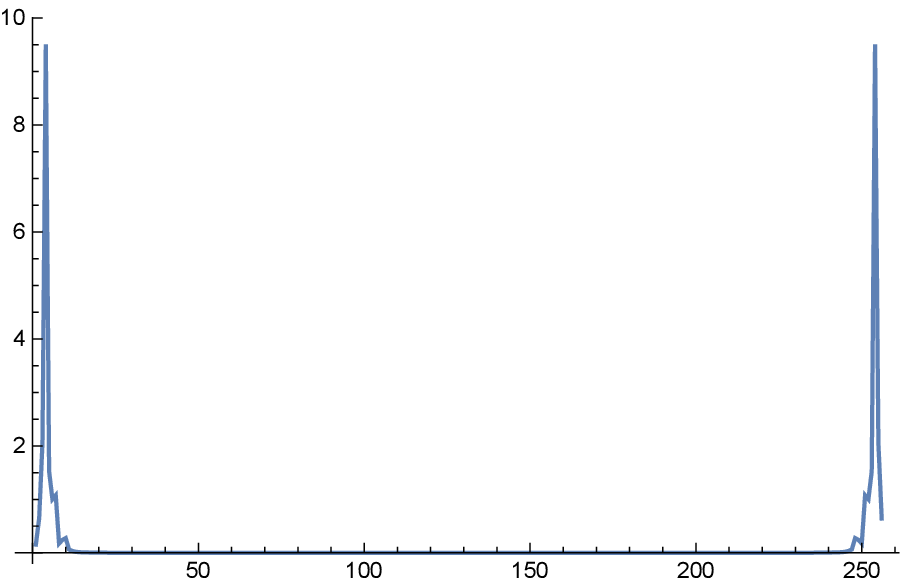}
\end{center}
\caption{Convolution of $V_{123}^+$ and $V_{123}^-$ for  $\omega\tau=3\pi/10$. Absolute values of the numerical Fourier transform are plotted. }
\label{fig:14} 
\end{figure}
\begin{figure}[htb]
\begin{center}
\includegraphics[width=5cm,angle=0,clip]{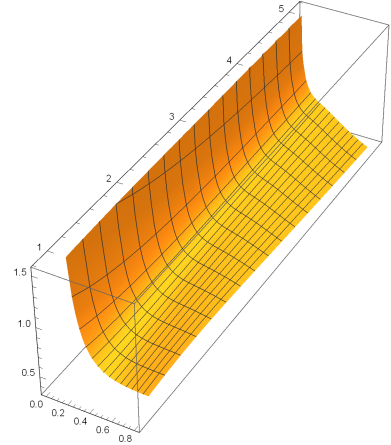} 
\end{center}
\caption{The surface $V_+$ given by the eq. (43) as a function of $N$ that runs $[0,0.8]$ and $z$ that runs $[0,1.5]$.  }
\label{fig:15}
\end{figure}
\begin{figure}[htb]
\begin{center}
\includegraphics[width=6cm,angle=0,clip]{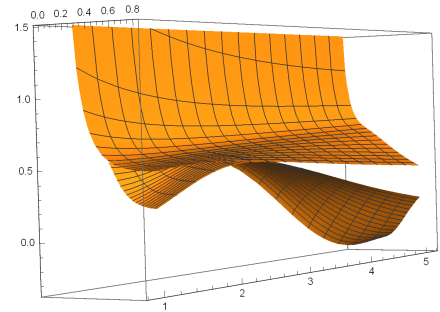}
\end{center}
\caption{The surface $V_+$ given by the eq. (43) as a function of $N$ and $z$ and the surface of  $V_+$ given by eq. (44), that is tangential to the surface of $V_+$ given by the eq. (43) at its peak.  }
\label{fig:16} 
\end{figure}

Fig.10 is the soliton wave function $V(z,N,\omega\tau)$(sum of $n=1,2,3$) with $N=0.5$ and $\omega\tau=\pi/2$. The wave function for $\tau=1-z$ and $\tau=t+z$ coincide.
 
 Fig.11 and Fig.12 are same as Fig.10 except $\omega\tau$ is chosen to be $\pi/10$ for the former and $3\pi/10$  
 for the latter.
 
 Fig.13 and Fig.14 are the absolute value of the convoluition of forward propagating soliton and backward propagating soliton. $\omega\tau=0$ for the former and $\omega\tau=3\pi/10$ for thre latter.
 
 Fig.15 represents the dependence on $N$ and $z$ of wave function $V(z,N,\omega\tau)$.
 
 The reduced velocity $V=V^{(1)}+V^{(2)}$ are parametrized as\cite{LR84}
\begin{equation}
V=B_1\sin\psi+B_2\sin 2\psi-A_2\cos 2\psi, \label{eq:V}
\end{equation}
whose maximum is assumed to occur at $\psi=\pi/2+\delta$, where $\delta$ is a small quantity, and within error of $\delta^2$, 
\begin{equation}
V_+=B_1+A_2+4B_2^2/(B_1+4 A_2),\label{eq:Vp1}
\end{equation}
and the value of the peak satisfies
\begin{eqnarray}
V_+&=&B_1\sin(\frac{A_2}{B_1+2B_2}+zV_+)\nonumber\\
&&+B_2\sin[2(\frac{A_2}{B_1+2B_2}+z V_+)]\nonumber\\
&&-A_2 \cos[2 (\frac{A_2}{B_1+2B_2}+zV_+)].\label{eq:Vp2}
\end{eqnarray}

Dependence of $B_1, B_2, A_2$ on $N$ and $z$ are given in \cite{LR84}.

 Fig.16 shows that a particular soliton solution which satisfy eq. (43) and eq.(44) can be obtained at a tangential point  of two planes which fixes $(N,z)$.

In Fig.\ref{fig:17},  $V_+$ of eq. (43) as a funcfion of $n$ and $z$ are plotted, and in the Fig.\ref{fig:18} , the right hand side of eq. (44) with $V_+$ as a funcion of $n$ and $z$, which is tangential to the surface of $V_+$ of eq. (43) are presented. ($0\leq N\leq 0.8$ , $1\leq z\leq 5$). A unique solution in the $(N,z)$ plane will be obtained by choosing $l_{mm}$ such that $\omega\tau$ becomes $0$ or $\pi/2$ and searching the tangential point of the two planes.

It is necessary to fix the position of the point where tangential plane of two surfaces become locally identical depending on directions of the phonon beams. 

\subsection{The convolution of $(2+1)D$ real wave functions and the Atiyah-Patodi-Singer index} 
The variables of a KhZa soliton is complex $z$, but wave function is ${\bf R}$.
In a simple situation in which a receiver is on the average height of the transducers which emits the KhZa soliton and the TR-KhZa soliton, the strength of the convolution of the KhZa soliton and the TR-KhZa soliton can be calculated by the $2D$ convolution program which exists in the library ASL of a supercomputer at RCNP. 

The Fig.\ref{fig:18} is the 2D real convolution of KhZa solitons, assuming a receiver is on a plane that passes the middle of ordinary wave source and TR wave source, and the distance between the point and the receiver is $z$. $N$ is a parameter that represents the ratio of the distance between the position of shock formation and the diffraction length. They are normalized as $0 < z<1$ and $0<N<1$.
 
\begin{figure}
\begin{center}
\includegraphics[width=8cm,angle=0,clip]{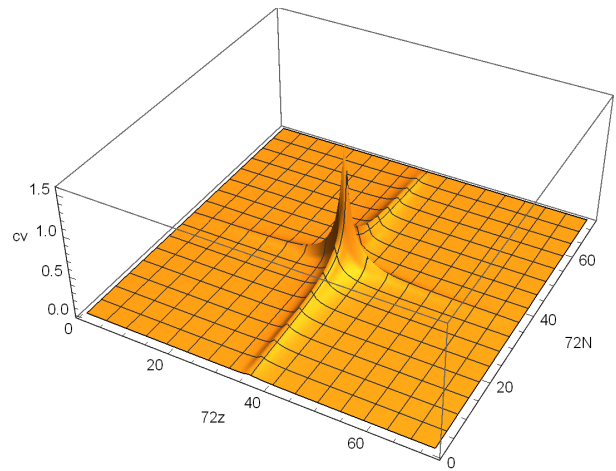} 
\caption{2D convolution of KhZa ordinary and TR solitons as a function of $z\in [0,1]$  and $N$. $72\times 72$ lattice points. }
\label{fig:17}
\end{center}
\end{figure}

\begin{figure}
\begin{center}
\includegraphics[width=8cm,angle=0,clip]{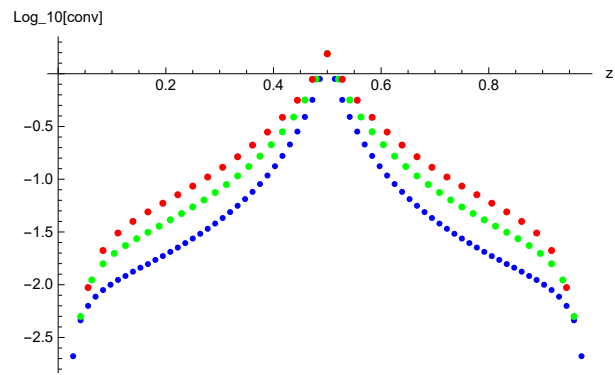} 
\end{center}
\caption{Logarithm of $2D$ convolution of KhZa ordinary and TR solitons $\log_{10}[conv]$ at $N=0.5$ as a function of $z$.  The number of mesh points are 36(red), 48(green) and 72(blue). }
\label{fig:18}
\end{figure}

\begin{figure}
\begin{center}
\includegraphics[width=8cm,angle=0,clip]{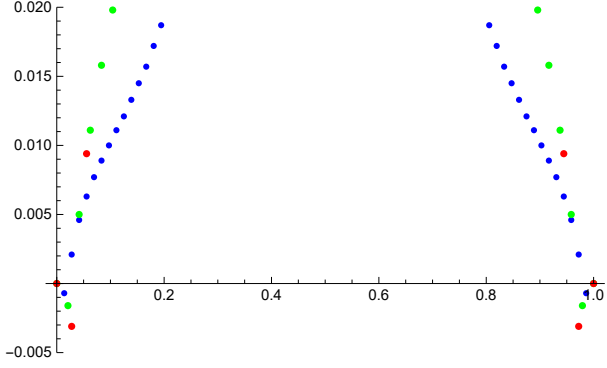} 
\end{center}
\caption{ $2D$ convolution of KhZa ordinary and TR solitons as a function of $z$ and at $N=0.5$  The number of mesh points are 36(red), 48(green) and 72(blue). }
\label{fig:19}
\end{figure}

 The $2D$ consists of the distance along the beam $z$ and the parameter $N$ in the equation (43). We choose lattice points $N_z=N_N=$18, 24 and 36. We compare cases of $dx=0.5$ and $0.6$ in proper units.  The parameter $R$ is chosen to be 1,  phase $\theta$ is chosen to be $0$ and the series of the Bessel function is truncated at $n=3$ in the Figs. \ref{fig:17}, \ref{fig:18} and \ref{fig:19}.

When $dx$ is large, the distance between transducer and receiver is large and the peak is reduced, The short ridge of the convolution value is along the $z$ axis, which means that the variation along $z$ axis is steeper than that along $N$ axis.
 The following figure is the logarithm $2D$ convolution on a $36\times 36$ lattice(red), $48\times 48$ lattice(green) and $72\times 72$ lattice(blue)  at $N=0.5$, as a function of $z$.  There appears local regions where logarithm of convolution is almost linear.
 
  Near $z=0$ and 1, there appears a point where the $2D$ convolution of KhZa solitons is negative. These singular points are dropped in the Fig.\ref{fig:18}. The Fig.\ref{fig:19} shows the singular point tends to approach the $z$ axis as the mesh size increases.
  
A $2D$ convolution as a function of $z$ at $N=0.5$ has a region where its logarithm is locally linear, and a point where it is negative. Near the boundary $Log_{10}(-0.007)=-2.1549+1.36438\sqrt{-1}$ has the positive imaginary part. It means that $t>0$ region is stable and $t<0$ region is unstable, and $n_+-n_-=2$, if gravitational anomaly is absent. 
 
Schwartz\cite{Schwartz61} defined set of all distribution in ${\bf R}^2$ as ${\mathcal D}({\bf R}^2)$ and distribution in ${\bf R}^2$
with compact support as ${\mathcal E}({\bf R}^2)$.
The convolution of $2D$ real distributions $u_1(x,y)\in {\mathcal D}'({\bf R}^2)$ and $u_2(x,y)\in {\mathcal E}({\bf R}^2)$ 
which is denoted as $u_1*u_2$ is numerically calculated as
\begin{eqnarray}
&&p(k_x,k_y)=\sum_{i_x=0}^{m_x-1}\sum_{i_y=0}^{m_y-1}u_1(i_x,i_y)u_2(k_x-i_x,k_y-i_y)\nonumber\\
&&=\sum_{j_x=0}^{m_x-1}\sum_{j_y=0}^{m_y-1}u_2(j_x,j_y)u_1(k_x-j_x,k_y-j_y),
\end{eqnarray}
where $k_x=0,\cdots,m_x-1$ and $k_y=0,\cdots, m_y-1$.

The Fourier transform of $u_1(x,y)$ and $u_2(x,y)$ are denoted as 
\begin{eqnarray}
\hat u_1(\xi,\zeta)=\int dx\int dy\, u_1(x,y)e^{\sqrt{-1}\langle x,\xi\rangle}e^{\sqrt{-1}\langle y,\zeta\rangle} ,  \nonumber\\
\hat u_2(\xi,\zeta)=\int dx\int dy\, u_2(x,y)e^{\sqrt{-1}\langle x,\xi\rangle}e^{\sqrt{-1}\langle y,\zeta\rangle}.
\end{eqnarray}
The symbol $\langle x, \xi\rangle$ means for a function $\phi\in C^\infty({\bf R}^2)$ and coordinates $x, \xi$, 
\begin{equation}
\langle x, \xi\rangle=x(\xi)=x(\phi \xi)=(\xi x)(\phi)=(\xi x)(1),
\end{equation}
when $\phi$ is $1$ near the overlapping region of the support of $x$ and that of $\xi$\cite{Hoermander83a}.

The Fourier transform of $p(k_x,k_y)$ is the product of $\hat u_1(\xi,\zeta)$ and $\hat u_2(\xi,\zeta)$.

When $u_1$ and $u_2$ are analytic function of $z\in{\bf C}$, the Cauchy-Riemann system
\begin{equation}
\partial u/\partial \bar z_j=g_j,\quad (j=1,2),
\end{equation}
where $g_j$ satisfies the compatibility condition\cite{Hoermander83b}
\begin{equation}
\partial g_j/\partial \bar z_k-\partial g_k/\partial \bar z_j=0,
\end{equation}
should be constructed.

When there are no cracks, $u_k$ are proportional to the solution $V(z,t)$ and one can check that
$g_k=0$. The situation is the same for the TR solution $V(z,-t)$.

Hoermander\cite{Hoermander83a,Hoermander85} defined principal symbols $p(\gamma)$
that transforms the coordinate system from $E$ labeled by $x$ to $F$ labeled by $\xi(x)$, and $p(\gamma)$ is defined on the section of a 
cotangent bundle $T^*(X)$. 
On $T^*(X)$ one can define one form $\omega$ and at $\gamma\in T^*(X)$ choose a tangent vector $t$, such that
\begin{equation}
\langle t,\omega\rangle=\langle \pi_*t, \gamma\rangle.
\end{equation}

On $T^*(X)$ one can define the symplectic form $\sigma=d\omega$ which is expressed  by the standard coordinate system $x,\xi$ as
\begin{equation}
\sigma=\sum_{j=1}^n d\xi_j\wedge dx_j.
\end{equation}
For an element $A$ of general linear group and for $n$ dimensional manifold, $\sigma(A(\Xi))(A(X))=\sigma(\Xi)(X)$,  
defines the symplectic group $Sp(n)$, which can be identified with a subgroup of $Sp(n,{\bf C})$\cite{Chevalley46}.

At a receiver convolution of phonons from different tranducers need to be measured and
analyzed. Different beam directions can be expressed by mapping of different coordinate system $p$ 
and by different sound velocity $c$.  We define the local coodinate system $\kappa_1$ and $\kappa_2$ such that
\begin{equation}
\kappa_2 \kappa_1^{-1}:\kappa_1(X_{\kappa_1}\cap  X_{\kappa_2})\to \kappa_2(X_{\kappa_1}\cap X_{\kappa_2}),
\end{equation}
where $X_{\kappa_1}$ and $X_{\kappa_2}$ are open sets in $X$.

In $X_{\kappa_1}\cap  X_{\kappa_2}$, $u=u_{\kappa_1} \circ \kappa_1^{-1}=u_{\kappa_2}\circ \kappa_2$.
When $f=du$ is an $n-$form on $X$ and coordinates are $\kappa_1(x)=(x_1,\cdots,x_n)$ and a $C^\infty$ map
 $\psi_1:X \to Y$, one can define $\psi_1=\kappa_1\circ\kappa_2$ and $\psi_1^* f$,
\begin{equation}
(\kappa_1^{-1})^* f=f_{\kappa_1} dx_1\wedge \cdots\wedge dx_n.
\end{equation}
Using a map $\psi_2:Y\to X$,
\begin{equation}
f_{\kappa_2}=(det \psi_2)\psi_1^* f_{\kappa_1},
\end{equation}
in $\kappa_2(X_{\kappa_1}\cap X_{\kappa_2})$.

When $\kappa_1$ system is defined as the beam line $T_3\to R_3$ in Fig. 5 as the real axis of the complex plane $S_1$, 
and $\kappa_2$ system is defined as the beam line $T_4\to R_3$ in Fig. 5 as the real axis of the complex plane $S_2$,
a Jacobian becomes dependent on the angle between the two real axis.

Soliton wave functions are disturbed by singularities and change their structure.

\section{Quaternion Neural Network and Topological Properties}
\label{sect:3}
In convolutional neural networks\cite{Aggarwal18}, input and output layers are defined on a $2D$ plane $L_q \times B_q$, where $L_q$ denotes the height of the wall and $B_q$ denotes the width of the wall, where the suffix $q$ indicates the depth of the $q$th layer. The system contains hidden layers having $F_q\times F_q\times d_q$ parameters for filtering. 
\begin{equation}
L_{q+1}=L_q-F_q+1,\quad B_{q+1}=B_q-F_q+1.
\end{equation}
For simplicity, we choose $F_q=3$ as in \cite{Aggarwal18}. 

We assume that the phonon can be approximated by KhZa solitons which is holomorphic on ${\bf R}^2$ when
\begin{equation}
1-2 b Nz+(1+b^2)n^2 z^2\ne 0.
\end{equation}
The condition can be checked by the initial condition.

The $p$-th filter in the $q$-th layer has parameters 
$W^{(p,q)}=[w^{(p,q)}_{ij 3}]$, $(i,j)=(1,2)$, and $q$-th layer are parametrized by $2D$ tensor $[h^{(q)}_{ij 3}]$.
Convolutional operations from $q$th layer to the $(q+1)$th layer are defined as
\begin{equation}
h^{(q+1)}_{ij3}=\sum_{r=1}^{F_q}\sum_{s=1}^{F_q} w^{(p,q)}_{rs3} h^{q}_{i+r-1\,j+s-1\,3},
\end{equation}
$\forall i\in\{1,\cdots,L_q-F_q+1\},\, \forall j\in\{1,\cdots, B_q-F_q+1\}$.

Each transducers and receivers exchange information on quaternion bases $t,x,y,z$. 

 Transducers $T_1,\cdots,T_M$ and receivers $R_1,\cdots,R_M$ are connected by weight function $w_{ij}$ $1\leq i \leq M, \, 1\leq M$. By padding, $L_q$ and $B_q$ increase from $M$ to $M+F_q-1$.

If the path length is taken $Mod[2 l_{mm}/c]$, i.e. back scattering is ignored, the equation (30)  becomes
 \begin{equation}
 hq_1-q_2 h=\left(\begin{array}{cc}
 0& -2\sqrt{-1}Z\\
 2\sqrt{-1}Z&0\end{array}\right),
 \end{equation}
 $q_1$becomes equivalent to $q_2$, by choosing $\tau=t\pm l_{mm}/c$, the matrix becomes $0_{\bf H}$.  Since we imposed no back scattering condition, 
 \begin{equation}
 -l_{mm}/c=l_{mm}/c, Mod.[2l_{mm}/c].
 \end{equation}
 
 In calculations of the loss function, it is necessary to take into account that the soliton beam line is defined in a finite region of $(\tau,{\bf z}_\perp)$ plane, where ${\bf z}_\perp$ is the $2D$ plane perpendicular to the beam direction. 
 
 Conformal wave functions in finite regions and symmetry protected topological bosonic phase of matter was discussed by Witten\cite{Witten16a}.  He proposed the symmetry protected topological (SPT) bosonic phase, which is disturbed by anomalies. In the three-manifold $X={\bf R}\times M$, where ${\bf R}$ is the parameter of time, there is Chern-Simons coupling of fields $A$:
 \begin{equation}
 CS(A)=\frac{e^2}{4\pi}\int_Y d^3 x \epsilon^{ijk}A_i \partial_j A_k, \quad k\in{\bf Z}.
 \end{equation}

Topological insurators and superconductors indicate that interaction of fermions and bosons are important.
Fermion topological phases are characterised by the index theory. 

Atiyah-Singer index theorem\cite{AS63} states that for an elliptic differential operator on compact manifold, the analytical index is equal to the topological index. The theorem was extended by Atiyah-Patodi-Singer (APS) \cite{APS75} to be applicable to an elliptic differential operators on manifolds with boundary.  As in Fig. 4, we resticted the variable $z=\tau{\bf I}+x{\bf i}+y{\bf j}$ to be on a complex plane with finite boundaries. In the process of arbitrary padding near the boundary, analytical continuation of the KhZa soliton wave function inside the area before padding of Fig. 4 to whole area after padding such that the boundary value becomes 0 will become possible

The APS index for the Dirac equation 
\begin{equation}
{\mathcal D}\Psi=\left(D_\mu \gamma^\mu-\left(\begin{array}{cc}0 & \sqrt{-1}m\\
                                                            -\sqrt{-1} m& 0\end{array}\right)\right) 
                                                            \left(\begin{array}{c}\psi_1\\
                                                                                     \psi_2\end{array}\right)=0,
\end{equation}
where $\psi_1(t,x_1,y_1)$ and $\psi_2(-t,x_2, y_2)$ are time reversed states that satisfy
\begin{equation}
T\psi(t,x_1,x_2)=\gamma_0\psi(-t,x_1,x_2),
\end{equation}
where $T$ is the time-reversal operator, is
\begin{equation}
Index({\mathcal D})=dim Ker({\mathcal D})-dim Coker({\mathcal D}).
\end{equation}

In order to characterize the index, one defines metric $g$ and gauge field $A$ and a spectral flow characterised by a parameter $s$;
\begin{equation}
(A_0,g_0)\to (A_0^\phi, g_0^\phi),
\end{equation}
such that $(A_s, g_s)$ coincides with $(A_0,g_0)$ at $s=0$ and $(A_0^\phi, g_0^\phi)$ at $s=1$.

When a system of Dirac fermions loses $T$ symmetry, and satisfies $(\sqrt{-1}\gamma^\nu D_\nu+\sqrt{-1}\mu)\psi=0$, the regularized partition function becomes
\begin{equation}
Z_{\psi,reg}=\prod_k \frac{\lambda_k}{\lambda_k+\sqrt{-1}\mu}.
\end{equation}

For large $\mu>0$, each eigenvalue $\lambda_k$ contributes $\sqrt{-1}$ or $-\sqrt{-1}$ to $Z_\psi$,
and
\begin{equation}
Z_\psi=|Z_\psi| exp(-\frac{\sqrt{-1}\pi}{2}\sum_k sign(\lambda_k)).
\end{equation}
Thus
\begin{equation}
Z_\psi=|Z_\psi|exp(\mp\sqrt{-1}\pi\eta/2)
\end{equation}
where $\eta=\lim_{s\to 0}\sum_k sign(\lambda_k)|\lambda_k|^{-s}$.

  The APS index theorem says that when boundary fermions give a $T$ conserving results.
  \begin{equation}
    exp(\mp \sqrt{-1} \pi\eta/2) exp(\pm\sqrt{-1}\pi (P-\hat A(R))=(-1)^{\mathcal I}),
  \end{equation}
 where $P$ is the instanton number, $\hat A(R)=\int_X \hat A(R) dx$ is the gravitational correction due to spacial curvatures.

  The formulae are valid also for Majorana fermions, and the relation
  \begin{equation}
  \frac{\eta}{2}=\frac{CS(A)}{2\pi}-2\frac{CS_{grav}}{2\pi}, \quad mod \,{\bf Z}
  \end{equation}
  was proposed. Gravitational anomalies are written in \cite{GSW87}.
 
 In 2+1 dimensional massless fermions $\Psi$, 
 \begin{equation}
 Z_\Psi=|Pf({\mathcal D})| exp(-\pi\sqrt{-1}\eta_R /2)
 \end{equation}
 where $Pf$ is the Paffian \cite{Witten16b}.
 
 The heat equation and the index theorem is discussed in \cite{ABP73,ABP75}.
 
  Optimum values of input values of hidden layers $F_q\times F_q\times d_q$ in the padding area would be defined by solving the APS boundary problem\cite{APS75,KM05a,KM05b,Witten16a,YWX17}.

  We replace $A_1, A_2$ by $Re[V(z,r,T)], Im[V(z,r,T)]$, and seek solutions which minimizes $CS(A)$ and effectively $V(z,r,T)$ becomes 0 at the boundary of the padded area and the square of differences of phonons beaming at receivers and TR-phonons beaming at the same receiver becomes minimum.
  
  We consider a $2D$ linear space $V$ and the exterior algebra $\wedge V$ and the Clifford algebra of $Q({\bf x})=Q(x_1{\bf e}_1+x_2{\bf e}_2)=x_1^2+x_2^2$ where ${\bf e}_1$ is the unit vector along the ordinary beam direction, and $Q({\bf y})=Q(\bar y_1 \bar {\bf e}_1+\bar y_2 \bar{\bf e}_2)=\bar y_1^2+\bar y_2^2$, where $\bar{\bf e}_1$ is the unit vector along the TR beam.   

The symmetric bilinear form associated with $Q$ and $V$ is
\begin{equation}
\langle x_1\wedge x_2, \bar y_1\wedge \bar y_2\rangle=det\langle x_i,\bar y_j\rangle=\left| \begin{array}{cc}
\langle x_1,\bar y_1\rangle & \langle x_1, \bar y_2\rangle\nonumber\\
\langle x_2, \bar y_1\rangle & \langle x_2, \bar y_2\rangle \end{array}
\right|.
\end{equation}

When a vector in the rectangular zone is defined as $\xi_1{\bf i}+\xi_2{\bf j}$ and the angle between  $\bf i$ and ${\bf e}_1$ is $\alpha$ and that between ${\bf i}$ and $\bar{\bf e}_1$ is $\beta$,
$det \langle x_i,\bar y_j\rangle$ becomes
\begin{eqnarray}
&&\left| \begin{array}{cc}
 x_1\bar y_1\cos\alpha\cos\beta  &  0 \\
0 & x_2 \bar y_2\cos\alpha\cos\beta \end{array}
\right|\nonumber\\
&&= x_1\bar y_1 x_2\bar y_2 \cos^2\alpha\cos^2\beta.
\end{eqnarray}

On a $2+1$ dimensional manifold with two spatial orientations, the metric $g_{ij}$ is related to the Dirac matrices $\gamma_i=\pm \sigma_i$ by
\begin{equation}
\gamma_i \gamma_j=g_{ij}\pm \sqrt{-1}\epsilon_{ijk}\gamma_k,
\end{equation} 
where $\epsilon_{ijk}$ is an antisymmetric tensor.

\section{Altland-Zirnbauer $DIII$ class in $(2+1)D$ spacetime}
\index{sect:4}
Altland and Zirnbauer \cite{AZ97} classified normalconducting mesoscopic systems in contact with superconductors and insulators into ten classes. In a Hamiltonian formulated by Bogoliubov and de Gennes (BdG), symmetry properties are specified by the group $G$, its representation $U$ and projection $P$ and the coset space $G/H$ 
\begin{equation}
U_X P_E U_X^{-1}=P_{{E}X^{-1}},
\end{equation}
where $E$ is a Borel set in $G/H$\cite{Mackey68}. Let $N$ be a commutative group of $G$, and $\hat N$ be the set of all characters on $N$. Let $N$ be translation in Eucledian space and ${\bf \alpha}$ be a rotation.
 For each character $\chi\in \hat N$, semidirect product $G=H\ltimes N$ with respect to $\bf \alpha$ is defined.

BdG hamiltonians are expressed as
\begin{equation}
\hat H=\sum_{\alpha,\beta}(h_{\alpha\beta}c_\alpha^\dagger c_\beta+\frac{1}{2}\Delta_{\alpha\beta}c_\alpha^\dagger c_\beta^\dagger+\frac{1}{2}\Delta_{\alpha\beta}^* c_\alpha c_\beta),
\end{equation}
where $c_\alpha^\dagger$ and $c_\beta$ are creation and annihilation operator in second quantization respectively and $h_{\alpha\beta}$ is a hermitian operator.
 
Haldane\cite{Haldane04} studied properties of $3D$ fermi surface and the effect of Berry curvature as a background effect.

Schnyder et al\cite{SRFL08} classified the symmetry operations on BdG hamiltonians into two types:
\begin{eqnarray}
&P:& H=-PHP^{-1}, PP^\dagger =1, P^2=1,\\
&C:& H=\epsilon_c C H^T C^{-1}, CC^\dagger=1, C^T=\eta_c C.
\end{eqnarray}
and showed that the BdG hamiltonian
\begin{equation}
{\mathcal H}(k)=\left(\begin{array}{cc}
                                     m & k\cdot\sigma(\sqrt{-1}\sigma_y)\\
                                     (-\sqrt{-1}\sigma_y)k\cdot\sigma& -m\end{array}
                   \right)
 \end{equation}
has 6 classes ($D, DIII$, $A, AIII$, $C, CI$).
                   
                   The $DIII$ class preserves TR symmetry but violates $SU(2)$ spin rotation symmetry.
The eigen state is a superposition of $(p+\sqrt{-1}p)$ and $(p-\sqrt{-1}p)$ waves.
Physical space of Dirac electron is classified by
\begin{eqnarray}
\Omega_1&=&\psi^\dagger \gamma_0 \psi,\nonumber\\
J^\mu&=&\psi^\dagger\gamma_0\gamma^\mu\psi,\nonumber\\
S^{\mu\nu}&=&\psi^\dagger\gamma_0\sqrt{-1}\gamma^{\mu\nu}\psi,\nonumber\\
K^\mu&=&\psi^\dagger \gamma_0\sqrt{-1}\gamma^{0123}\gamma_\mu\psi,\nonumber\\
\Omega_2&=&\psi^\dagger\gamma_0 \gamma^{0123}\psi,
\end{eqnarray}
and classified fermion spinors into 6 types\cite{Lounesto93,Lounesto01}.  

The 5th class is characterized by $\Omega_1=\Omega_2={\bf K}=0$, ${\bf S}\ne 0$ is called Majorana spinor.
The 6th class is characterized by $\Omega_1=\Omega_2={\bf S}=0$, ${\bf K}\ne 0$ is called Weyl spinor.

Ryu et al\cite{RML12} studied electric and thermal response of topological insurators, and in symmetry class $DIII$ of space-time dimension $d=(2+1)$, predicted the gravitational anomaly. 
\section{Quaternion Fourier Transform }

Fourier transformation of complex functions in the topological vector space \cite{Horvath66} was established by Schwartz \cite{Schwartz61}. 
In his theory, the Bochner-Minlos theorem\cite{Wiki20} plays an essential role. It says that if there is a nuclear space $A$, a characteristic functional $C$ and for any $z_j\in {\bf C}$ and $x_j\in A$,
\begin{equation}
\sum_{j=1}^n\sum_{k=1}^n z_j \bar z_k C(x_j-x_k)\ge 0
\end{equation}
is satisfied, there exists a unique measure $\mu$ and the dual space $A'$, 
\begin{equation}
C(y)=\int_{A'} e^{{\sqrt -1}\langle x,y\rangle} d\mu(x).
\end{equation}

In engineering, quaternion functions apperar in color image processings \cite{Berthier13} and recurrent neural networks\cite{PRMLTDMB19}. 

The two-sided quaternionic Fourier transform (QFT) was introduced by Hitzer and Sangwine\cite{HS13,Hitzer22} extending the pioneering work of Ell \cite{Ell13}. Georgiev et al.\cite{GMKS13,MGS14} considered, due to noncommutativity of quaternions, a left-sided QFT, a right-sided QFT and a two-sided QFT, using the finite integral.

In $(2+1)D$ subspaces
\begin{equation}
T=x_1+x_2{\bf i}+x_3{\bf j}=\left[ \begin{array}{cc}x_1+x_2\sqrt{-1}& x_3\\
                                  - x_3&x_1-\sqrt{-1}x_2\end{array}\right],
\end{equation}
and its pair
\begin{equation}
\bar T=x_1+x_2{\bf i}+x_3{\bf j}=\left[ \begin{array}{cc}x_1-x_2\sqrt{-1}& x_3\\
                                  - x_3&x_1+\sqrt{-1}x_2\end{array}\right],
\end{equation}
both $\in M_2({\bf H})$ satisfies 
\begin{equation}
T \bar T=\left[\begin{array}{cc} x_1^2+x_2^2& x_3^2\\
                                                       x_3^2& x_1^2+x_2^2\end{array}\right].
\end{equation}
It means that the gaussian structure remains after quaternion Fourier transformations.

\section{Discussion and perspective}
From the theoretical point of view, TR-NEWS concepts search singularity on the border of cone of propagating sound by convolutions of regular and time reversed waves. We want to maximize the convolution of the output from original signal
 ${\mathcal G}_\varphi f(\omega_1,\omega_2;b_1,b_2)$ and output from the TR signal
 $\widehat{ {\mathcal G}_\varphi f}(\omega_1,\omega_2;{\hat b}_1,{\hat b}_2)$ as small as possible.  That is minimize
\begin{equation}
\int_{{\bf C}^2}[{\mathcal G}_\varphi f(\omega_1,\omega_2; b_1,b_2)-\widehat{ {\mathcal G}_\varphi f}(\omega_1\omega_2;\hat b_1,\hat b_2)]^2 d\omega_1 d\omega_2
\end{equation}
by proper choices of $b_1,b_2,\hat b_1,\hat b_2$.

 $(2+1)D$ quaternion representation of instant form can be transformed to that of front form, and time steps can be selected as $\tau_1,\cdots,\tau_K$.

Chua showed that in memristic circuits, output frequency shows Devil's staircase structure\cite{Chua11,Chua18} that there are stable output frequency regions. Each steps may correspond to emergences of equivalent quaternion wave functions in the projective space.
We showed a possible method of applying the quaternion neural network to NDT.
In order to suggests a practical applicationable concepyts , it is necessary to optimize the phonetic pulse shape and minimize the difference of convolutions of original wave and that of TR wave.

 For optimization of getting positions of cracks in a rectangular media, quaternion neural network is a promissing method. Effects of quantum gravity through metrics in gauge theories could be analyzed using the same approach as suggested by Gan \cite{Gan21}.
 
 The choice of quaternion projective space on $2D$ planes is expected to reduce number of training parameters. Numerical calculation using the Generalized Conjugate Residual (GCR) method proposed by Luescher\cite{Luescher03} applied to Weyl fermion systems is under investigation. 
 
 Via analysis of solitonic phonons which is TR invariant, and the APS index measurement using neural network technique is expected to provide information of the gravitational anomaly.
\begin{itemize}
\item The Lie-Trotter formula and parametrization of time by $\tau=t\pm \frac{l_{mm}}{c}$, and extension of regular functions in the cone area due to Hille-Yosida theory are importance for achieving the convergence of evolution equations.
\item The choice of quaternion projective space on $2D$ planes is expected to reduce number of training parameters, which needs further studies.
\end{itemize}
A program for GPU, or vector processors running under MPI, which is inspired by AI, and search parameters that reproduce patterns of phonons emitted from transducers on a wall, scattered by cracks inside materials, and detected by receivers on a wall on the other side, are under investigation.

\begin{acknowledgements}
S.F. thanks Prof. Stan Brodsky and Prof. Guy de T\'eramond for valuable informations on conformal field theory. Thanks are also due to the RCNP of Osaka University for allowing checking FFT programs using its super computer, and Tokyo Institute of Technology for consulting references and a guidance of supercomputer programmings.
\end{acknowledgements}

%
%



{\small

\end{document}